# Supramolecular transmission of soliton-encoded bit streams over astronomical distances

**Authors:** W. He[1], M. Pang[1]*, D. H. Yeh[1], J. Huang[1], C. R. Menyuk[1,2], P. St.J. Russell[1].

**Affiliations:**

[1]Max Planck Institute for the Science of Light, Staudtstrasse 2, 91058 Erlangen, Germany.

[2]Department of Computer Science and Electrical Engineering, University of Maryland Baltimore County, Baltimore, Maryland 21250, USA.

*Correspondence to: meng.pang@mpl.mpg.de

**Abstract:** A stream of optical pulses, transmitted over long distances in optical fiber, will be affected by a variety of noise sources, leading to degradation in the signal-to-noise ratio. This noise accumulation sets generic capacity limits of all fiber-based optical signal transmission systems and has long been regarded as unavoidable. We report that by tailoring long-range, non-covalent inter-pulse interactions, optical solitons in a fiber laser loop can robustly couple to each other and self-assemble into supramolecular structures that exhibit long-term stability, elementary diversity and the possibility of information encoding. We demonstrate error-free transmission of such self-assembled solitonic structures over many astronomical units without any active retiming, opening up the possibility of using bit-bit interactions to overcome noise accumulation in optical fiber telecommunications and bit-storage systems.

**One Sentence Summary:** Non-covalent interactions between optical pulses allow error-free transmission of supramolecular soliton streams over astronomical distances.

**Main Text:** The second law of thermodynamics states that the disorder (entropy) of the universe increases with time (*1*). In optical-fiber-based telecommunications and bit-storage rings (*2–6*), the transmitted signal follows a similar law: both additive and multiplicative noise accumulate with transmission distance (*7, 8*). In addition, long-range interactions between optical pulses in these systems have proven almost impossible to control (*9, 10*), generally leading to a gradual degradation of the optical encoded information and imposing fundamental limitations on system performance (*7*). An analogy may be drawn with self-assembly in natural systems—a ubiquitous phenomenon in which elementary components join together via non-covalent (i.e., weakly bound) interactions to form stably-ordered, macroscopic structures. The most dramatic example is DNA (*11*) in which strands of nucleotides are bound together by weak forces. Carrying biological information, DNA is able to stably transmit genetic codes from one generation to the next.

    Inspired by the concept of "self-assembly" from supramolecular biochemistry (*11, 12*), we introduce here transmission of supramolecular bit streams of optical pulses (see Fig. 1A). In contrast to previous work in which strong (covalent) interactions are employed to form soliton molecules (*13–16*), here we demonstrate that weak interactions between solitons in specially designed fiber laser loops can allow the self-assembly of large population of solitons into highly-ordered structures (*12*). Due to the long-range nature of the weak forces, the soliton spacing in these supramolecular structures can range from tens to hundreds of picoseconds, facilitating real-time monitoring using fast electronics. Moreover, these supramolecular solitonic structures exhibit remarkable long-term stability—we have observed stable storage of pulse patterns in this fiber

laser loop over many days with no need of any electronic retiming, corresponding to error-free transmission of a soliton stream over distances of many astronomical units (*5*, *8*). At the same time, these supramolecular structures exhibit considerable flexibility—we were able to change the soliton spacing and the total soliton number, and add or remove individual solitons without damaging the underlying supramolecular structures.

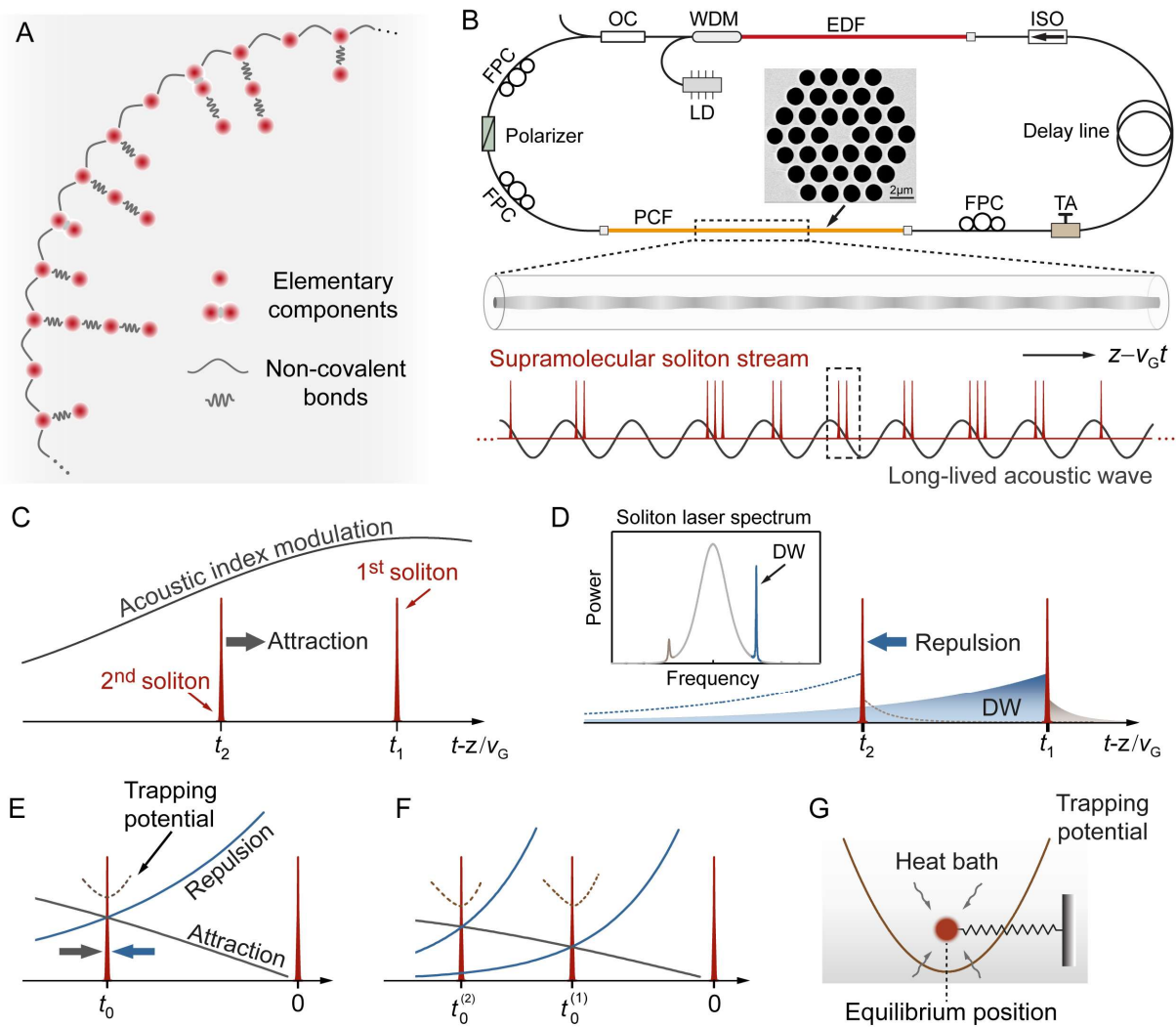

**Fig. 1. Principle of supramolecular assembly of optical solitons**. (**A**) Conceptual sketch of a self-assembled supramolecule, consisting of different elementary components bound together by weak (non-covalent) interactions. (**B**) Sketch of the experimental set-up, with an inset scanning electron micrograph (SEM) of the photonic crystal fiber. The supramolecular soliton stream propagating in this fiber laser loop drives an acoustic resonance in a PCF core, creating an "optomechanical lattice". EDF: erbium-doped fiber; WDM: wavelength-division multiplexer; LD: laser diode, OC: output coupler; FPC: fiber polarization controller; TA: tunable attenuator; ISO: isolator. (**C**) Within each unit of the optomechanical lattice, a long-range "force of attraction" between the solitons arises due to the optomechanical effect. (**D**) A competing "force of repulsion" between the two solitons appears due to dispersive wave perturbations. The inset figure shows a typical soliton spectrum with asymmetric Kelly sidebands. (**E**) Competition between these two forces forms a temporal potential, trapping the 2nd soliton. (**F**) Stable multi-soliton units can form through the cascaded build-up of trapping potentials. (**G**) The timing jitter of an individual soliton in a supramolecule is analogous to the Brownian motion of a particle trapped in a harmonic potential.

laser loop over many days with no need of any electronic retiming, corresponding to error-free transmission of a soliton stream over distances of many astronomical units (*5*, *8*). At the same time, these supramolecular structures exhibit considerable flexibility—we were able to change the soliton spacing and the total soliton number, and add or remove individual solitons without damaging the underlying supramolecular structures.

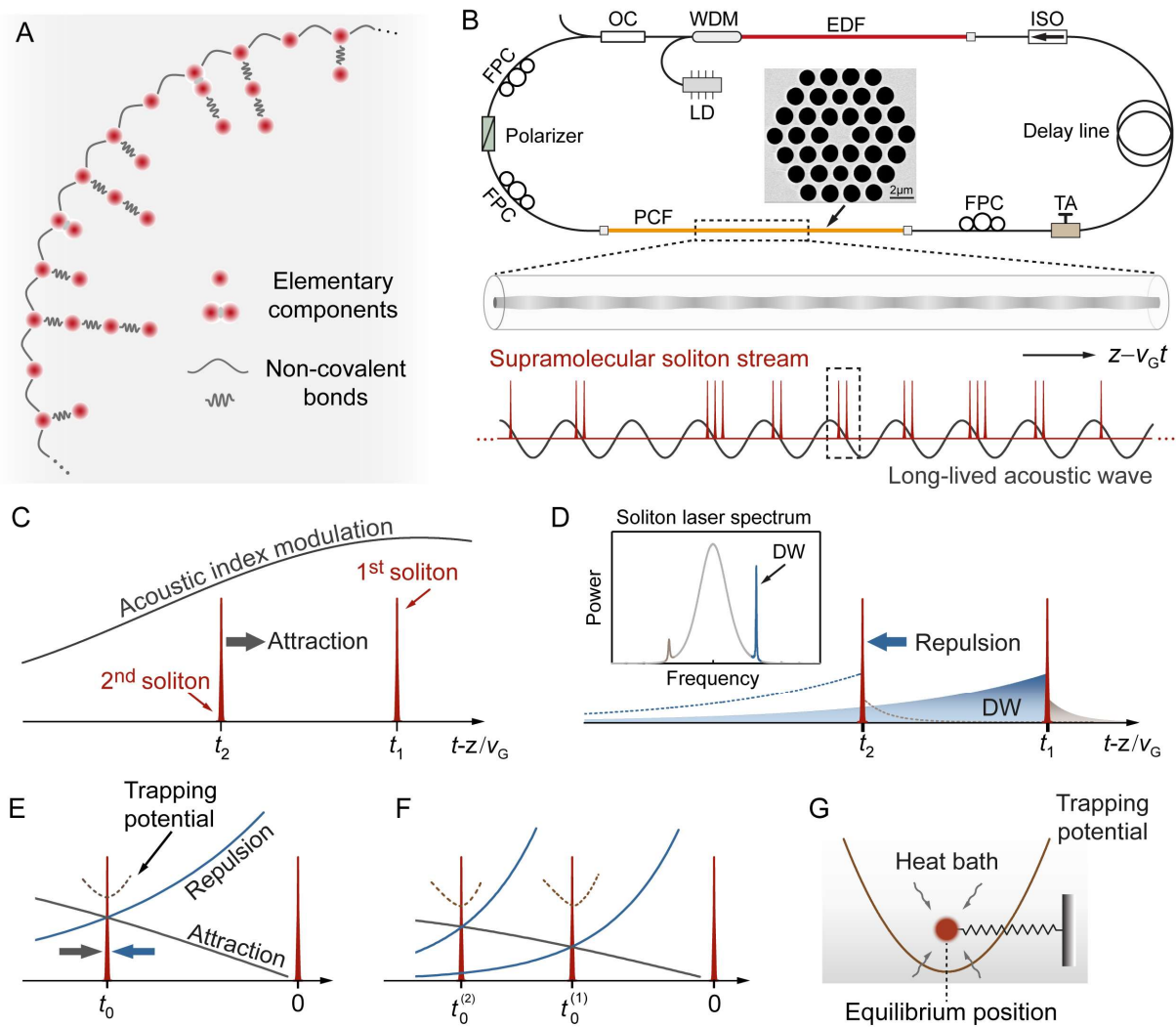

**Fig. 1. Principle of supramolecular assembly of optical solitons**. (**A**) Conceptual sketch of a self-assembled supramolecule, consisting of different elementary components bound together by weak (non-covalent) interactions. (**B**) Sketch of the experimental set-up, with an inset scanning electron micrograph (SEM) of the photonic crystal fiber. The supramolecular soliton stream propagating in this fiber laser loop drives an acoustic resonance in a PCF core, creating an "optomechanical lattice". EDF: erbium-doped fiber; WDM: wavelength-division multiplexer; LD: laser diode, OC: output coupler; FPC: fiber polarization controller; TA: tunable attenuator; ISO: isolator. (**C**) Within each unit of the optomechanical lattice, a long-range "force of attraction" between the solitons arises due to the optomechanical effect. (**D**) A competing "force of repulsion" between the two solitons appears due to dispersive wave perturbations. The inset figure shows a typical soliton spectrum with asymmetric Kelly sidebands. (**E**) Competition between these two forces forms a temporal potential, trapping the 2nd soliton. (**F**) Stable multi-soliton units can form through the cascaded build-up of trapping potentials. (**G**) The timing jitter of an individual soliton in a supramolecule is analogous to the Brownian motion of a particle trapped in a harmonic potential.

The mode-locked fiber laser loop we employed to demonstrate transmission of soliton supramolecules is sketched in Fig. 1B (details in supplementary section S1). A 2-m-long solid-core silica photonic crystal fiber (PCF) with an acoustic core resonance at 1.887 GHz (*17*) was inserted into the laser cavity. Upon increasing the pump power of the Er-doped fiber amplifier (EDFA), a variety of different supramolecular soliton streams were generated in the laser cavity, all of which globally ordered by optomechanical interactions (*18–20*). A long-lived acoustic core resonance is driven coherently by the soliton stream, and acts back on the pulses, linking them together by modulating their carrier frequencies, and forming a "temporal optomechanical lattice" with a period equal to that of the acoustic vibration (*18*). This global lattice divides the soliton stream into many units of identical length, within which multiple solitons can settle (see Fig. 1B), analogously to the periodic sugar-phosphate backbone in a DNA molecule (*11*).

Long-range binding of multiple solitons in each optomechanical unit originates from the balance between attractive and repulsive inter-soliton forces. As the solitons "ride" on the acoustic wave whose envelope co-propagates at the same group velocity (*18*), the index modulation caused by the acoustic wave causes the soliton carrier frequency to shift while propagating (see Fig. 1C). The amplitude of this frequency shift is determined by the slope of the underlying index modulation. Since the two solitons within one acoustic period are located at different positions (see Fig. 1C), their frequency shift rates differ. This divergence of soliton frequencies, acting in concert with the group-velocity dispersion of the optical fiber, causes an effective force of attraction (*21*). A competing long-range repulsive force arises from long-lived dispersive waves that are shed by the solitons during propagation. Dispersive wave emission, originating from the periodic perturbation of solitons (*22, 23*), is a generic phenomenon in both optical soliton communications systems and soliton fiber lasers (*5, 16, 24*). As illustrated in Fig. 1D, the dispersive wave corresponding to the first higher-frequency sideband shed from the 1st soliton would propagate faster than the soliton and eventually reaches the 2nd soliton, perturbing it through cross-phase modulation (*25*). Such dispersive wave perturbations effectively create a repulsive force between the two solitons (*26*), balancing the force of attraction due to the optoacoustic effect (*21*). In principle, more than two solitons can be stably bound within one unit (see Fig. 1B) through the cascaded balance of long-range forces (see Figs. 1, F and G). The entire supramolecular structure then consists of a chain of units, each containing its own quota of bound solitons—not unlike the individual nucleotides in a DNA molecule weakly held together by hydrogen bonds.

The cooperative balance of different long-range forces produces stable soliton spacing, effectively creating inter-solitonic "springs" that trap the solitons at each equilibrium position (see Fig. 1, E-G). As a result, noise-induced fluctuations in soliton-soliton spacing (timing jitter) can be viewed as analogous to the Brownian motion of a particle trapped within a harmonic potential (*21*) as depicted Fig. 1E), which can be modeled using the Langevin equations (*27, 28*). The mean square deviation of the pulse timing jitter can then be expressed as $<z(t) - z_0>^2 = N_{\Delta\omega}/(2\Gamma KB)$, where $z(t)$ is the time-dependent pulse spacing, $N_{\Delta\omega}$ the noise amplitude, $\Gamma$ the damping parameter, $K$ the optomechanical stiffness, and $B$ describes the fiber dispersion (*21*). When pulses are trapped in a harmonic potential, the timing jitter no longer grows with time, even though the "heat bath" (noise source) continuously disturbs the system.

A typical supramolecular soliton stream with a chain of units, each containing 0, 1, 2 or 3 trapped solitons, was recorded using a fast detector and an oscilloscope. The resulting time-domain trace is shown in Fig. 2A, where the underlying grid is globally locked to the 1.887 GHz acoustic resonance in the PCF core (with a period of 532 ps). The duration of individual solitons was measured to be 650 fs. This self-assembled soliton stream, once formed, was found to be indefinitely stable, the pulse spacing being 80 ps between the 1$^{st}$ and 2$^{nd}$ solitons and 70 ps between the 2$^{nd}$ and 3$^{rd}$ in each unit. In the experiments we monitored the supramolecular soliton stream over 1000 min (see Fig. 2B), and observed no obvious degradation in signal-to-noise ratio. This corresponds to error-free transmission of the soliton stream over twelve billion kilometers (~84 astronomical units) in a freely-running fiber laser loop (See more details in Supplementary Section S2).

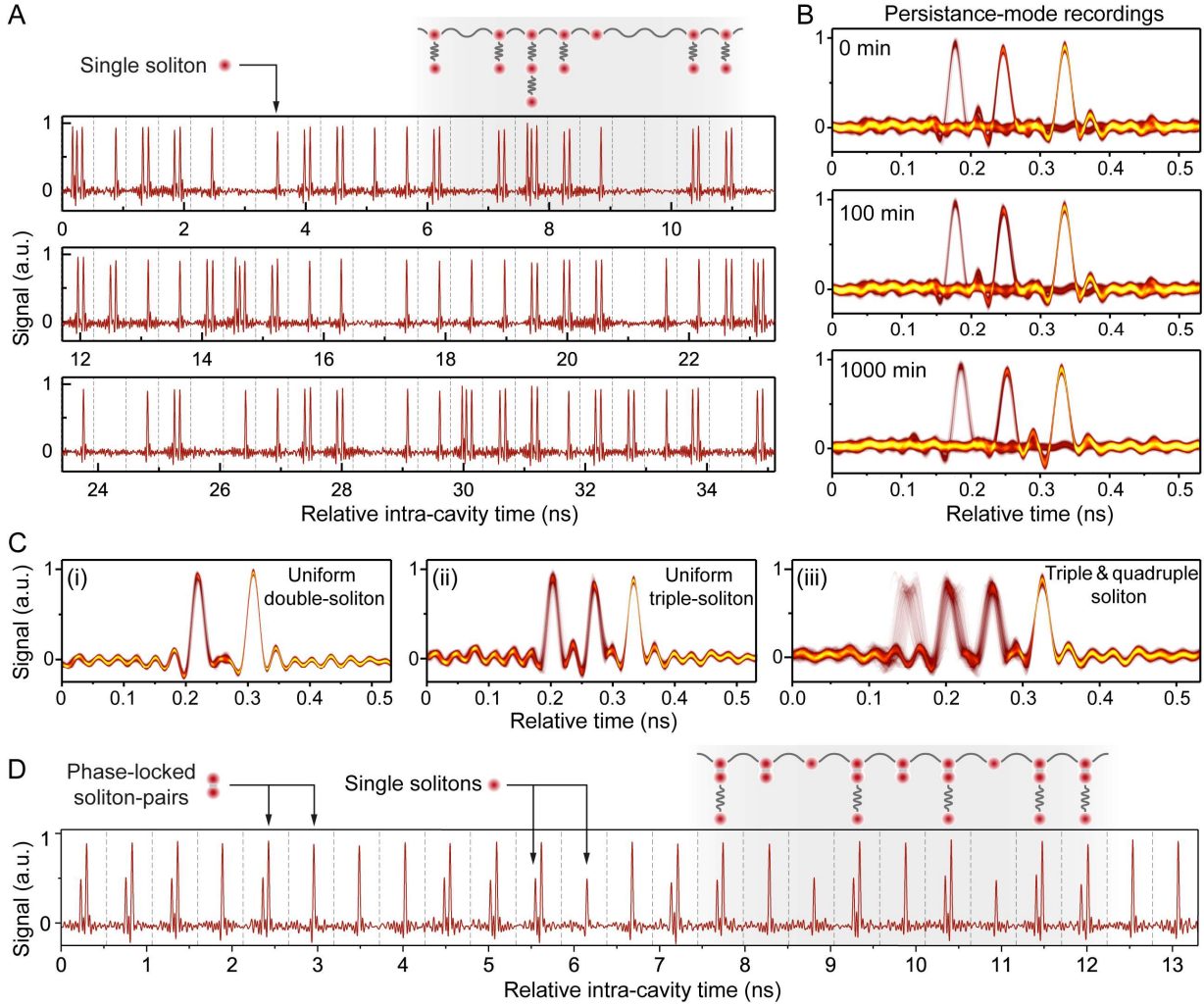

**Fig. 2. Long-term stability and configurational diversity of the supramolecular soliton stream.** (**A**) The time-domain trace of a typical supramolecular soliton stream (only 66 out of 154 time-slots are shown). (**B**) Persistence-mode recordings of this supramolecular soliton stream (similar to the eye-pattern diagram) at 0 min, 100 min, and 1000 min. (**C**) Persistence-mode recordings of three typical supramolecular soliton streams containing (i) uniform double-soliton units, (ii) uniform triple-soliton units, and (iii) both triple- and quadruple-soliton units. (**D**) Time-domain trace of a typical supramolecular soliton stream containing both single-solitons and phase-locked soliton-pairs as fundamental building-blocks.

The maximum number of solitons that can be trapped in each supramolecular unit is limited mainly by the ratio between the acoustic period and the internal soliton spacing, and the total number of units is determined by the acoustic resonant frequency and the cavity length (round-trip time). The highest number of units reached in the experiments was more than one thousand. If both the cavity length and the PCF structure are kept constant, the maximum EDFA gain will limit in practice the maximum number of solitons that can be self-assembled, which could amount to several thousand. In the experiments we could only partially control the fine structure of the supramolecular soliton stream (see supplementary section S3). For example, we were able to reproducibly generate uniform structures with single (*18–20*), double, or triple soliton units (Fig. 2C) throughout the entire structure. We could also generate a supramolecular soliton stream with both triple and quadruple soliton units (Fig. 2C). In addition, the elementary diversity of such supramolecular structures can be greatly expanded by incorporating more types of fundamental building-blocks. For example, we have observed that both single solitons and phase-locked soliton pairs (*13–15*) can be incorporated as building-blocks in the supramolecular structure; the time-domain trace of such a structure is recorded in Fig. 2D. This type of supramolecular structure is constructed from both non-covalent (weak) and covalent (strong) bonds, which further consolidates the analogy with biochemistry (*12*). Note that, due to the low temporal resolution (~20 ps) of the oscilloscope, in Fig. 2D we can merely observe that some pulses in this supramolecular stream have amplitudes twice higher than the others. We measured the real-time spectrum of this soliton stream using the time-stretch dispersive Fourier transform (TS-DFT) method (*29–31*). The results show that these higher-amplitude pulses in the time-domain trace correspond to phase-locked soliton pairs with a much shorter pulse spacing of ~4.5 ps (see supplementary section S4).

The weak nature of the non-covalent interactions makes the soliton streams highly reconfigurable. For example, their inner structure can change in response to variations in the non-covalent forces (see Fig. 3, A and B). We have found that the spacing between the solitons in one unit can be continuously tuned to and fro over a large range, while maintaining the overall supramolecular structure. By placing a tunable attenuator in the laser cavity (see Fig. 1A) we were able to adjust the cavity gain and loss, permitting continuous tuning of the dispersive wave intensity (see Fig. 3C). In the experiments we were able to increase its intensity by 200% (see Fig. 3D), dramatically reinforcing the repulsive force between the solitons (see Fig. 3B). As shown in Fig. 3D, this led to an increase in the internal soliton spacing from 40 ps and 116 ps in an all-double-soliton structure. We were also able to cycle the soliton spacing back and forth by adjusting the cavity length so as to vary the amplitude of the acoustic wave and thus the attractive force (see more details in Supplementary Section S5).

The supramolecular pattern could also be changed by abruptly perturbing the system. Both adding and removing individual solitons (see Fig. 3, E and G) are possible. Two examples are shown in Fig. 3, F and H. While a sudden jump in EDFA pump power resulted in generation of additional solitons in the supramolecular stream as recorded in Fig. 3F, a sudden drop in the pump power resulted in some solitons dropping out as recorded in Fig. 3G (More details in Supplementary Section S6). Remarkably, the supramolecular soliton streams both before and after the pump power variations were stable, demonstrating the possibility of fine control (information encoding) of the supramolecular structure (*18*, *32*).

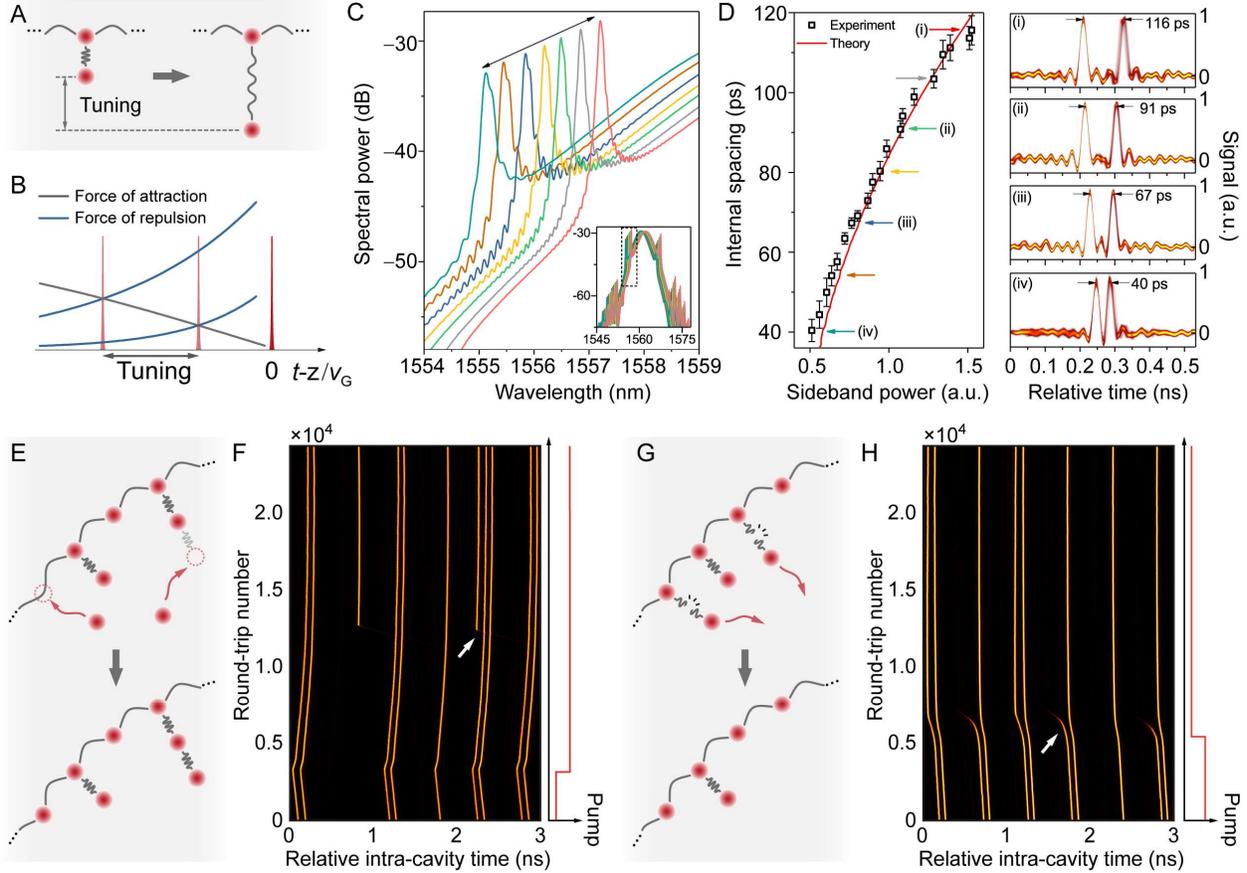

**Fig. 3. Structural reconfigurability of the supramolecular soliton stream.** (**A**) and (**B**) Tailoring the pulse spacing within each unit of the supramolecular structure by varying the non-covalent interactions. (**C**) Adjusting the intensity of the 1st higher-frequency sideband when the soliton bandwidth is invariant as shown in the inset. (**D**) Tuning the soliton spacing from 40 ps to 116 ps by varying the dispersive wave intensity. The experimental data is plotted as black squares, agreeing well with the red theoretical curve (see ref (*21*)). Four typical persistence-mode traces are shown on the left (i) – (iv). (**E**) and (**G**) Conceptual sketch of adding and removing solitons to and from existing supramolecular structures. (**F**) and (**H**) Experimental results demonstrating the dynamics of adding and removing individual solitons to and from the supramolecular structures.

In contrast to conventional soliton molecules and crystals (*16*, *29*, *32*), in which solitons in close proximity strongly interact through their tailing fields (covalent bonding), the formation of supramolecular soliton structures in this work relies on weak, long-range, non-covalent bonding. While, due to the long-range nature of the non-covalent forces, the soliton-soliton spacing can be hundreds of times longer than the duration of individual solitons, the supramolecular soliton streams are no longer localized structures of light composed of several phase-locked solitons. In fact, phase-locked covalent soliton molecules can form the building-blocks of supramolecular assemblies (see Fig. 2D and Supplementary Section S4). The self-assembled structures of solitons and soliton molecules are distributed over the entire fiber loop, with a large population of components and a built-in hierarchy. In contrast to chemical or biological supramolecules (*11*, *12*), which are generally static self-assemblies, the supramolecular soliton streams demonstrated here are essentially packets of optical pulses. Their traveling nature suggests analogies with the collective behavior of a colony of marching ants, which arises from interactions between many individuals, each following a simple set of rules, rather than from top-down instructions from the queen (*33*).

From the standpoint of nonlinear optics, solitonic supramolecules are solutions of dissipative nonlinear optical systems such as passively mode-locked fiber lasers (*16*, *34*, *35*), optical-bit-storage fiber loops (*4*, *6*) or optical fiber telecommunications systems with non-trivial nonlinearity levels (*2*, *5*, *7*). Being nonlinear attractors, these solutions are relatively immune to noise. The demonstration of self-assembled soliton streams reveals that a properly-designed nonlinear optical system can support many structurally-protected solutions (*29*, *36*) that can be used to transmit (store) optically-encoded information over unprecedented distances (durations).

**Funding:** Max Planck Society (MPG)

**Author contributions:** The concept was proposed by M.P., W.H., and P. St.J. R., the experiments were carried out by W.H., M.P., D.H.Y, and J.H., and the results were analyzed by W.H., M.P., and C.R.M. The paper was written by all authors.

**Competing interests:** Authors declare no competing interests.

**Supplementary Materials:**

Section S1-S6

Figures S1-S7

References (*37, 38*)

# Supplementary Materials for

## Supramolecular transmission of soliton-encoded bit streams over astronomical distances

W. He, M. Pang, D. H. Yeh, J. Huang, C. R. Menyuk, P. St.J. Russell

Correspondence to: meng.pang@mpl.mpg.de

**This PDF file includes:**

Supplementary Section S1 to S6
Figs. S1 to S7

# Contents

# S1. Experimental set-up

The platform we employed to perform the experiments was a soliton fiber laser with a ring configuration (*18-20*), as sketched in Fig. S1. The 2-m-long solid-core silica photonic crystal fiber (PCF) spliced into the laser cavity had a core diameter of 1.95 μm, with $R_{01}$ mechanical core resonance at 1.887 GHz (*17*). A tunable delay-line (TD) was used to adjust the cavity length, and a tunable attenuator (TA) to adjust the cavity loss. An optical isolator (ISO) ensured the unidirectional operation of the laser. The total cavity length was ~17 m, corresponding to a free spectral range (FSR) of ~11.7 MHz. The cavity average dispersion was about –0.046 ps$^2$/m, ensuring operation of this laser in the soliton regime (*36*). The cavity loss could be tuned from ~6 dB to ~ 30 dB by adjusting the intra-cavity tunable attenuator (TA) with a resolution of 0.02 dB. The laser cavity length could be varied using an intra-cavity tunable delay line (TD) with a tuning range of 0.15 m and a resolution of 3 μm.

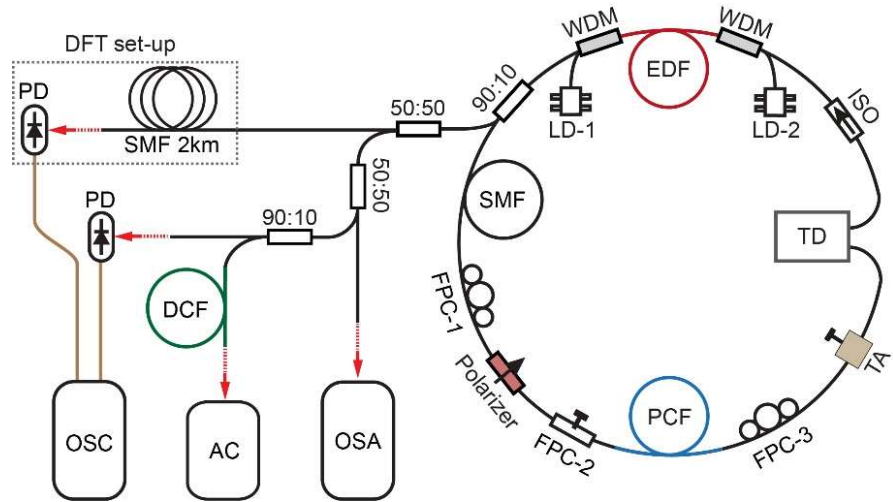

**Fig. S1. Experimental set-up.** A unidirectional soliton fiber laser cavity and the laser output was recorded using a fast oscilloscope (OSC) and a high-resolution optical spectrum analyzer (OSA). An autocorrelator (AC) is used to measure the output pulse duration. EDF: Erbium-doped fiber; WDM: wavelength division multiplexer; LD: laser diode; SMF: single-mode fiber; PC: polarization controller; PCF: photonic crystal fiber; TA: tunable attenuator; TD: tunable delay line; ISO: optical isolator; PD: photo detector; DCF: dispersion-compensation fiber.

When the laser was harmonically mode-locked by the optomechanical effect in the PCF (*18-20*), the cavity round-trip time was effectively divided into 154 time-slots by the optically-driven GHz-rate mechanical vibration in the PCF core, providing an effective optomechanical lattice. Within each time-slot (mechanical vibration cycle) of the lattice, one long-range multi-soliton unit can be trapped long-term. For steady operation of this mode-locked laser, a proper working point of the NPR needed to be set through fine adjustments of PC-1 and PC-3 in the laser cavity, while no further adjustment of the PCs is required for long-term preservation of the intra-cavity soliton supramolecules.

The 10% laser output from the 90:10 coupler was divided into 4 parts using three couplers as shown in Fig.S1, in order to be connected to different diagnostic devices. The output from one of the ports was detected using a 30-GHz photodetector and a 33-GHz oscilloscope (OSC), so as to obtain the time-domain trace or the laser output. The response time of this real-time detection is ~20 ps which sets the pulse width shown in all the plots that were recorded using the OSC. The timing jitter of the OSC in sampling is ~2 ps which gives the measurement error in reading fine structures of the supramolecules. For measuring the duration of laser solitons, a second-harmonic autocorrelator was used with a time resolution of 20 fs. The optical spectrum of the laser output was measured using an optical spectrum analyzer with a resolution of 0.01 nm. One part of the laser output was reserved for performing the time-stretched dispersive Fourier transform (TS-DFT) measurement (*29-31*), which was used for characterizing one of fundamental building-blocks of the supramolecular soliton stream: the phase-locked soliton-pair.

### S2. Long-term preservation of supramolecular soliton stream:

In the experiments we obtained a variety of hybrid soliton supramolecules consisting of soliton units with different amount of solitons (i.e. different long-range bound-states), which also exhibited perfect long-term stability. The full plot (over one complete cavity round-trip) of a typical hybrid supramolecular soliton stream is plotted in Fig. S2A (partially shown in Fig.2A), which includes null-soliton, single-soliton, double-soliton, and triple-soliton units. The entire pattern has been observed to be perfectly preserved after each cavity round-trip (as shown in Fig. S2B) and was able to last for more than one week without any obvious noise accumulation. In the experiments the supramolecule soliton stream demonstrated in Fig. S2A was recorded at 0 min, 100 min, and 1000 min after it was generated. The results (only the highlighted part in Fig. S2A is plotted due to the limited figure size) are shown in Fig. S2C, confirming its long-term preservation.

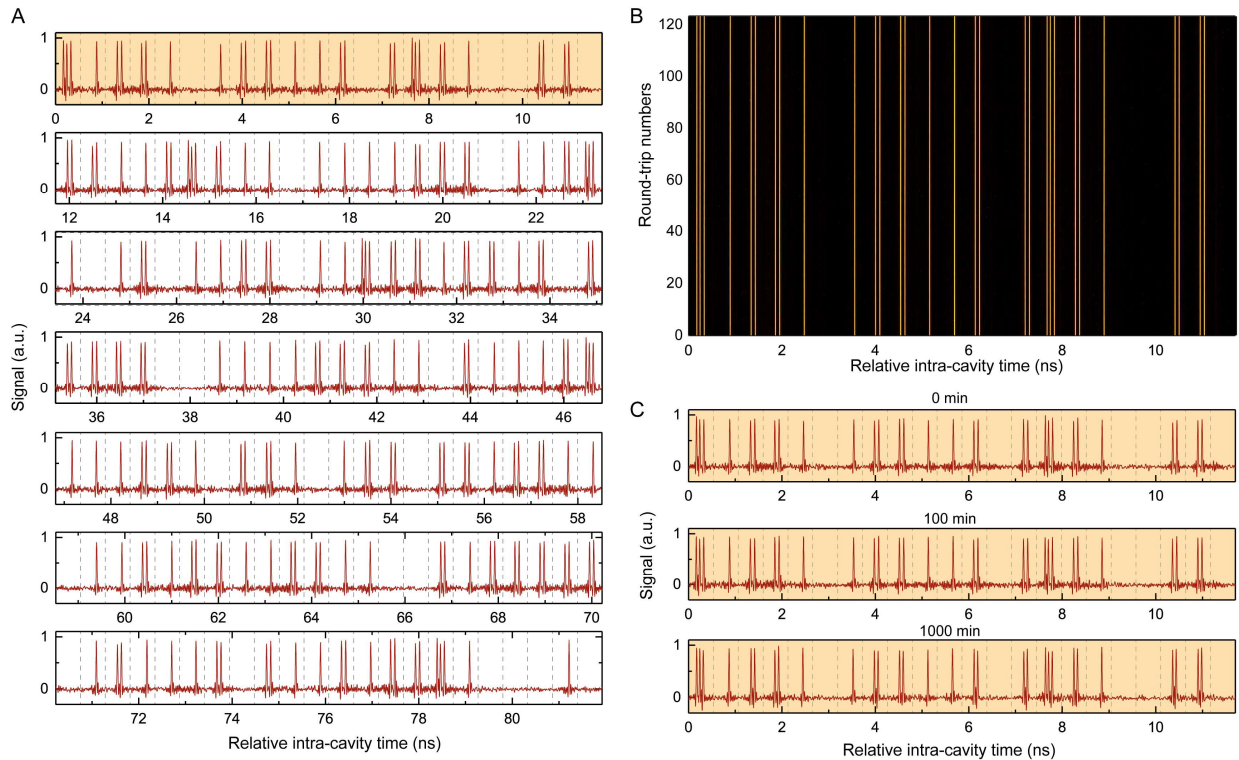

**Fig. S2.** (**A**) The complete supramolecular soliton stream over one cavity round-trip. (**B**) The round-trip plot of the same soliton supramolecule over 123 consecutive round-trips (~10 μs). (**C**) The highlighted part in (A) recorded at 0 min, 100 min, and 1000 min after the soliton stream was generated.

In Fig.2B, we can notice a slight change of internal spacing at 1000 min compared to the previous two records. This is actually due to the long-term mechanical drifting of the FPC in the cavity which leads to slight change of the working point (the balance point of gain and loss in the laser cavity, see ref (*18*)). Nevertheless, the supramolecular structure was able to constantly adjust itself according to these "environmental" changes while stably preserving the basic pattern.

In the experiments we could only partially control fine structures of the hybrid supramolecules. As shown in Fig. 3, E–H in the main text, we have experimentally demonstrated the possibilities of adding/removing some individual solitons through varying the laser pump power. It should be, however, possible to achieve fine control of the patterns of supramolecular structures, for example through further perfecting the previously-developed technique (*18*).

## S3. Uniform all-double-soliton supramolecules

In the experiments we were able to reproducibly generate supramolecular soliton streams with homogeneous patterns, including the all-double-soliton (ADS) and the all-triple-soliton (ATS) supramolecules. The typical recording of the ADS state is shown in Fig. S3 below.

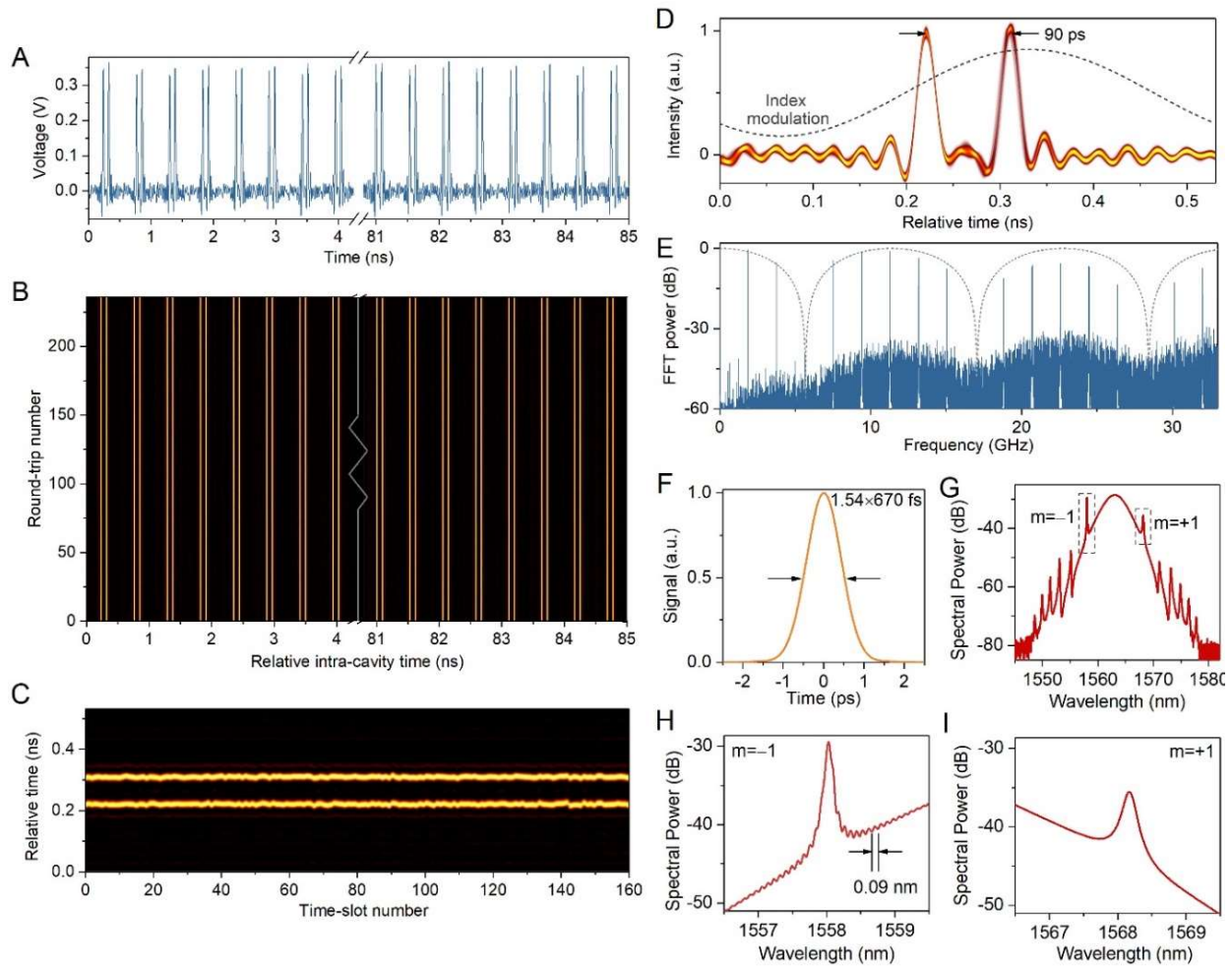

**Fig. S3. Illustrations of the all-double-soliton supramolecule.** (**A**) Time-domain trace of the supramolecule recorded using the OSC. Due to the limit of figure size, the trace is plotted over only 8 ns (the cavity round-trip time was ~85 ns). (**B**) A plot of a longer time-domain trace over 230 cavity round-trips in the persistence mode. (**C**) The time-domain traces in the 160 time-slots of the lattice plotted in parallel. The vertical axis corresponds to the relative time in each time-slot (the period of lattice was ~0.531 ns), while the horizontal axis corresponds to the 160 consecutive time-slots. (**D**) The time-domain trace of the ADS supramolecule recorded under persistence mode, exhibiting a stable 90-ps internal spacing. (**E**) The FFT power spectrum of the time-domain trace with a sinusoidal envelope (dashed-line). (**F**) The autocorrelation trace of the ADS supramolecule stream. (**G**) The optical spectrum of the ADS supramolecular stream, with the dominant sideband marked in dashed-box. (m=−1-order at shorter wavelength and m=+1-order at longer wavelength.) (**H**) Zoom-in of the spectrum in (G) at the dominant sideband (m=−1). (**I**) Zoom-in of the spectrum in (G) at the other sideband (m=+1-order).

Fig. S3A shows the ADS supramolecule recorded using the OSC over one cavity round-trip (with only the starting and ending parts of one round-trip stream). As we increased the recording time, we could plot its time-domain trace over many cavity round-trips (the round-trip time was ~85 ns) in a persistence way as shown in Fig. S3B, which exhibits a perfect preservation of those double-soliton units in the optomechanical lattice. We also illustrate the experimental data in a third way as shown in Fig. S3C, in which we separately plot the time-domain trace within each time-slot of the lattice (the period of the lattice was ~0.532 ns) along the ordinate, and we repeated such plot for the 160 consecutive time-slots along the abscissa. Such a plot can exhibit the relative positions of the double-soliton units in the optomechanical lattice. Fig. S3D shows the same pulse train recorded under the persistence mode over a single time-slot span, from which we can see the characteristic inter-soliton spacing of 90 ps which is the identical in all the time-slots. The FFT spectrum of the time-domain trace shown in Fig. S3E exhibits a sinusoidal envelop with a period of 11.1 GHz which agrees well with the 90 ps pulse spacing. The pulse duration measured using the autocorrelation trace is only 670 fs (See Fig. S3F), while the soliton-soliton spacing in each unit is two-orders of magnitude longer (90 ps). The optical spectrum of the ADS supramolecule stream, as shown in Fig. S3G, which at first glance has trivial difference compared to any typical soliton laser, featuring sech$^2$-shaped profile and the characteristic Kelly sidebands, which corresponds to dispersive waves at phase-matched wavelengths (*22, 23*). However, the spectral fringe that appeared only in the vicinity of the dominant sideband (m=−1 order) indicates the phase-locking between the corresponding dispersive wave and the soliton (as shown in Fig. S3H), while on the other side (the m=+1-order sideband) does not exhibit any interferometric fringe (as shown in Fig. S3I). The interferometric fringe in the vicinity of the m=−1 sideband exhibits a period of 0.09 nm (or 11.1 GHz), agreeing well with the 90 ps internal soliton spacing. All of these features are in accordance with our theoretical model described in (*21*). The all-triple-soliton case (ATS) was also obtained in our experiments has been observed to have similar properties compared to the ADS case (See ref (*21*)).

## S4. Phase-locked soliton pairs as building-blocks

Phase-locked soliton-pairs (or soliton molecules) that are bound via "covalent" interactions can also behavior as fundamental building-blocks in the supramolecular soliton stream. The example shown in the Fig. 2H demonstrates the case in which some units of the supramolecular stream were actually constituted by a single soliton and a phase-locked soliton pair. Due to the low temporal resolution of the detection system, the phase-locked soliton pairs in this time-domain trace (see Fig. S4, A and B) were shown as some pulses with amplitudes twice higher than the others. By using the dispersive Fourier transform (DFT) technique (*29-31*), we were able to resolve these phase-locked soliton pairs more clearly. In the experiments a 2-km-long SMF-28 was used to stretch the supramolecular soliton stream (see set-up in Fig. S1), and the measured DFT-signal was shown in Fig. S4C. We could reveal that each pulse with twice amplitude in Fig. S4A was stretched into a wide envelop with high-contrast fringes, which is the typical signature of a phase-locked soliton pair (*29*). In contrast, the single soliton with lower amplitude was stretched into a more relative weak envelop without interferometric fringes. Note that the DFT signal traces of all the soliton pairs features interferometric dips at the centers of the envelopes, indicating the fixed phase-difference of π between the two solitons in all the pairs (*14, 29*). These phase-locked soliton pairs, as fundamental building-blocks, can interact non-covalently with other building-blocks, forming a stable supramolecular structure.

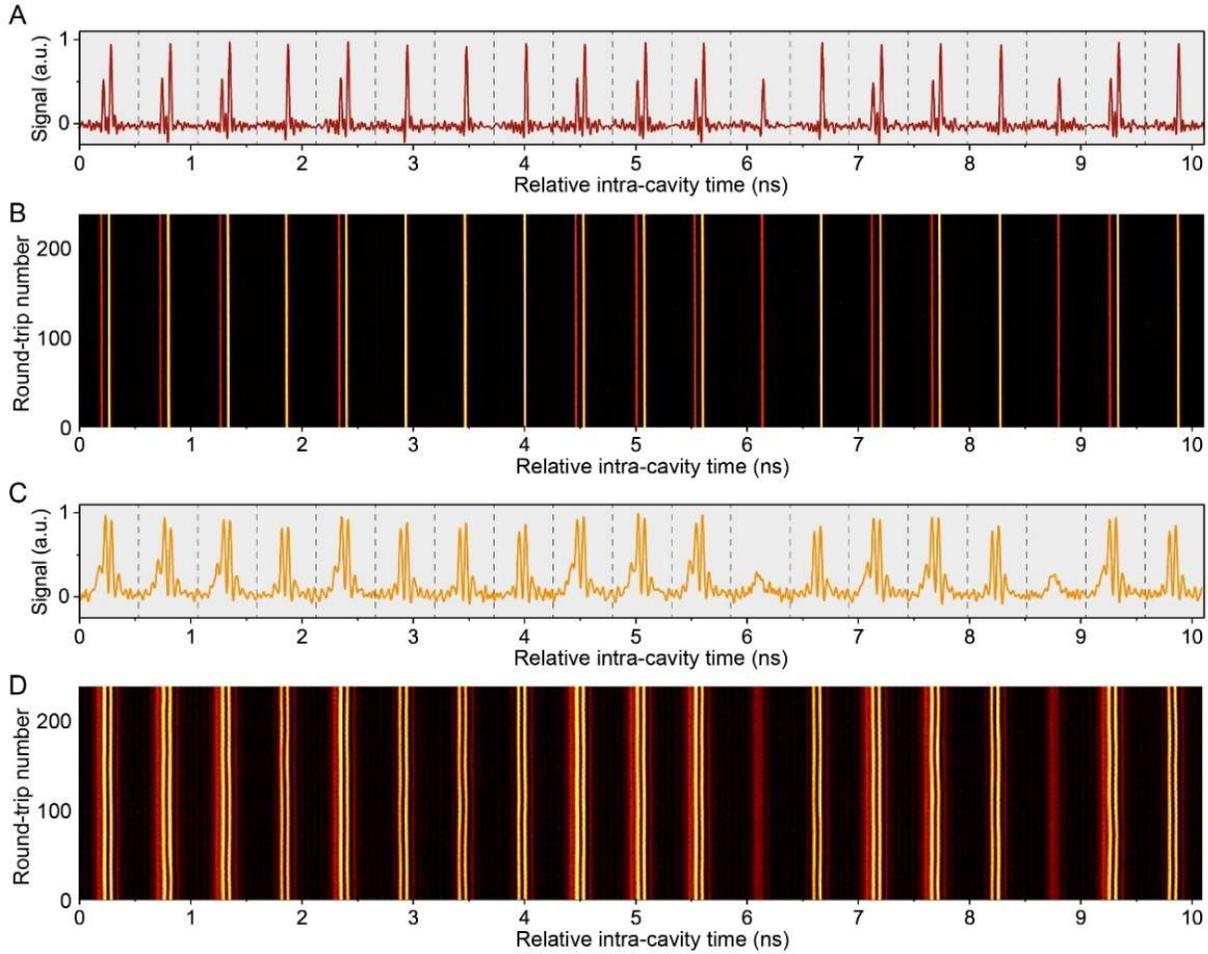

**Fig. S4.** (**A**) The supramolecular soliton stream that consists of both individual solitons and phase-locked soliton pairs (with twice as high amplitude). Only 19 out of 154 time-slots are plotted due to space limitation. (**B**) The round-trip plot of the same soliton stream over 246 round-trips. (**C**) The DFT-signal of the same soliton stream after temporal stretching over 2-km long SMF. (**D**) The round-trip plot of the DFT signal for over 246 round-trips.

Since a considerable amount of the fundamental building-blocks of this soliton supramolecule as shown in Fig. S5 are phase-locked soliton pairs, strong interferometric fringes can also be observed on the laser optical spectrum. As shown in Fig. S5A, the optical spectrum (recorded using the OSA) has a fringe period of ~1.85 nm. The autocorrelation trace shown in Fig. S5B features two side peaks which give a direct evidence of the soliton-pair structure with 4.51 ps soliton-soliton spacing. Note that unlike the DFT recording, the OSA and autocorrelator can only exhibit the time-averaged recording. Due to the co-existence of many single solitons, the fringe contrast on the optical spectrum is not every high (see Fig. S5A), and the amplitude of side peaks in the autocorrelation trace is lower than half of the central peak amplitude (see Fig. S5B).

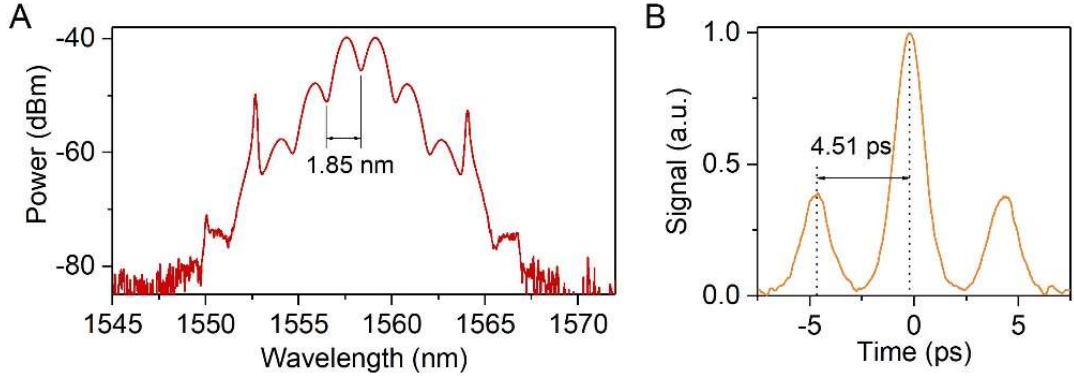

**Fig. S5.** (**A**) The optical spectrum of the soliton supramolecule demonstrated in **Error! Reference source not found.**, which includes both phase-locked soliton pairs and single solitons as fundamental building-blocks. (**B**) The autocorrelation trace of the soliton stream.

The elementary diversity we observed in the experiments can take far more appearances beyond the case demonstrated in the main text. For example, the soliton-pairs do not necessarily always take the first position within the units as shown in Fig. S4A, i.e., it could also be the reversed order, and both types of order can co-exist in the same supramolecule soliton stream. In such a more complicated case, the repulsive forces exert by a single soliton and a soliton-pair would become different, which leads to different internal spacing of the long-range bound-state solitons. Moreover, we have also observed that two bound-soliton pairs can also form a long-range bound-state, in which the soliton-pairs behave just like individual solitons in exerting the long-range forces described in ref (*21*). Consequently, due to the introduction of phase-locked soliton pairs, the simplest long-range bound-state of two fundamental building-blocks in one unit can take four different appearances (single-single, single-pair, pair-single, pair-pair), and they would have different internal spacing as we have observed from some recorded streams in which all of these 4 appearances co-exist. In addition, soliton-pairs with phase-difference other than π has also been observed in other soliton supramolecules, which further contribute to diversity of fundamental diversity.

## S5. Soliton spacing tuning: Supplementary experiments

The soliton spacing tuning of the ADS supramolecule can be achieved by tailoring both the intensity of dispersive wave and the amplitude of mechanical vibration in the PCF core. In the experiments we varied the intensity of the $m=-1$ order dispersive wave by adjusting the cavity loss, which was implemented by adjusting the in-cavity tunable attenuator inserted in the cavity (see Fig. S1). When the total cavity loss was gradually varied from ~6 dB to ~11 dB, we observed that the intensity of the dominant Kelly sideband dropped by around three times (as shown Fig. 3C and 3D in the main text). In this process, the average optical energy out of the EDFA decreased, however, by less than 5%, which could be compensated by slightly increasing the pump power of the EDFA. The almost unchanged intra-cavity optical energy ensured that the optomechanical effect in the PCF remained invariant. As illustrated in Fig.3, A–D, the increase in the dispersive wave would result in stronger force of repulsion between the two solitons, leading to a larger soliton spacing. In practice, while the basic structure of the ADS supramolecule was stably preserved in the laser cavity, the soliton spacing in the double-soliton unit could be tuned at will within a broad range from 40 ps to 116 ps. The maximum soliton spacing that we observed in the experiments was ~170 ps, achieved through increasing further the dispersive wave intensity, which however demanded a higher pump power.

The soliton spacing could also be tuned through varying the force of attraction, as illustrated in Fig. S6A. This was implemented by slightly adjusting the cavity length using the intra-cavity delay line (see Fig. S1). By gradually decreasing the cavity length, the free spectral range (FSR) of the cavity increased, corresponding to a tuning of the lattice frequency of the ADS supramolecule toward the mechanical resonance frequency of the PCF. In our experiment, as the lattice frequency was tuned from 1.8725 GHz to 1.882 GHz (Fig. S6B), the soliton spacing in the double-soliton unit decreased from 87 ps to 63 ps which was induced by an increase of the force of attraction as illustrated in Fig. S6C. This tendency rolled over when we further increased the lattice frequency around 1.883 GHz, agreeing well with the theoretical predictions in ref (*21*). Similarly, the soliton spacing in the triple-soliton unit could also be tuned in both ways.

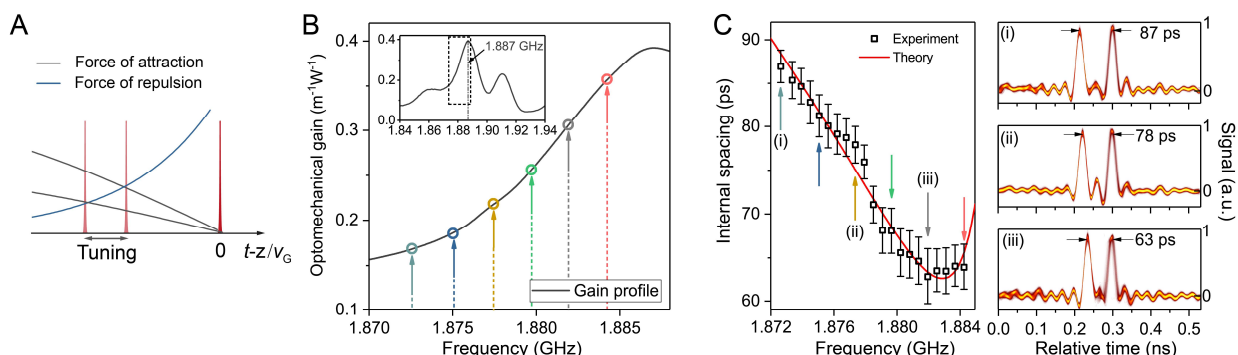

**Fig. S6. Soliton spacing tuning of the all-double-soliton supramolecule by tailoring the acoustic wave.** (**A**) Tailoring the double-soliton spacing by varying the force of attraction. (**B**) Tuning the lattice frequency of the ADS supramolecule by varying the cavity length. The optomechanical gain band of the solid-core PCF is plotted in the inset, showing a mechanical resonance at 1.887 GHz. (**C**) The soliton spacing first decreases as the lattice frequency of the supramolecule is tuned toward the mechanical resonance frequency, and then rolls over at ~1.882 GHz. The soliton spacing could be tuned from 63 ps to 87 ps, and three oscilloscope traces are shown on the left (i)–(iii). The experimental data is plotted as black squares, which agrees well with the red theory curves (see ref (*21*)).

For measuring the soliton spacing, long-time averaging was used to obtain its mean value, and the root mean square of the spacing jitter was used for the error bars in Fig. 3D, Fig. S6B. In the experiments, we observed slightly higher spacing jitter (with a maximum jitter value of ~5 ps) near the upper and lower edges of the tuning range. The large spacing jitter occurred at the edges of tuning because we always optimized our system near the midpoint of the tuning range. Nevertheless, if all the system parameters were maintained in the recording, we did not observe degradation of the spacing jitter over time.

## S6. Dynamic process of adding and removing solitons

By strongly perturbing the pump power, we were able to add or remove solitons to or from the existing supramolecular soliton streams. In this section, we provide some experimental details during such dynamic processes. In the case of adding solitons, as shown in Fig.3F, the pump power was modulated such that there is an abrupt increase of ~15% with a 5-ns rising edge. Such a modulation of laser pump power caused two consequences. Firstly, all the existing solitons would experience sudden decreases in their group velocities due to their increased intensities, which lead to longer cavity round-trip time and therefore the "bending" of the yellow lines at the time of the step of increased pump. Secondly, and more interestingly, the noise background of the laser arose and some noise spikes turned into solitons that have equal intensities of the previously existing solitons. These newly-appeared solitons were then trapped in the trapping potentials as described above, increasing the soliton numbers in some units (see the zoom-in of the transition process in Fig. S7A corresponding to the region marked by white arrow in Fig.3F). The entire supramolecular structure, nevertheless, remained stable during this process, and the birth of new solitons itself involves complicated gain- and nonlinearity-related dynamics (*37*).

We could also remove some individual solitons from some specific time-slots without destroying the entire supramolecular structure, as shown in Fig.3H. When the pump power is decreased by ~10% over a trailing edge of 5-ns, many of the double-soliton units degraded into single-soliton units, with the $2^{nd}$-soliton deviated and diminished over a few hundreds of cavity round-trips. The zoom-in of the unit highlighted by white arrow in Fig.3H was shown in Fig. S7B. The underlying mechanism of this soliton moving process could possibly be interpreted in the following way. Firstly, when pump power is decreased, the soliton peak intensity was reduced and the soliton acquired a larger group velocity. The second-soliton trapped in some units would tend to escape from the trapping potential. The net frequency-shift experienced by the $2^{nd}$-soliton would then lead to a lower gain, since its carrier frequency deviated from the gain maximum. As a feedback, the group velocity of the $2^{nd}$-soliton would become higher due to its lower intensity, leading to a faster escaping rate. Such positive feedback would accelerate this escaping process, and finally the $2^{nd}$-soliton would be diminished in the background.

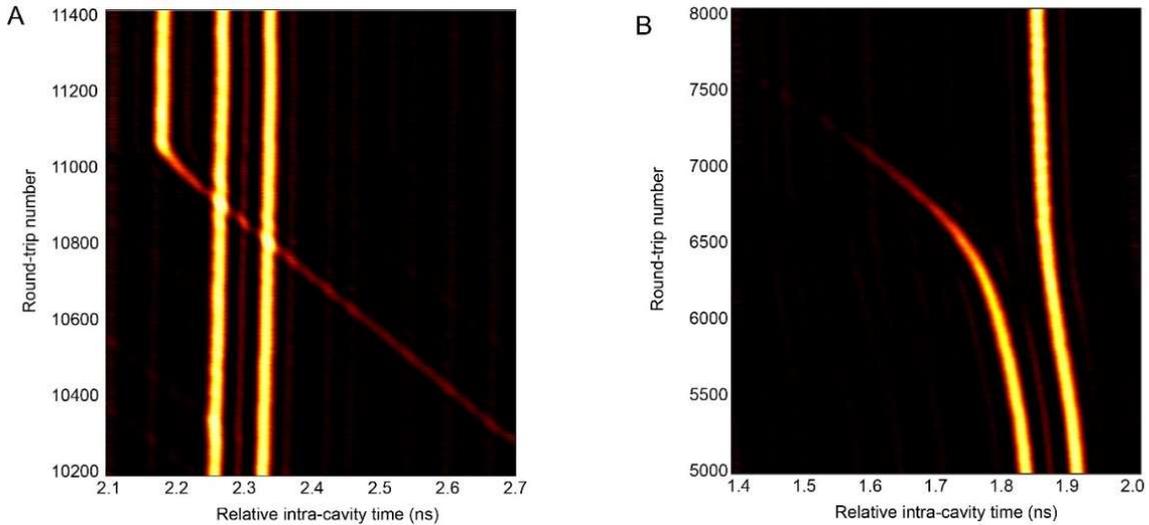

**Fig. S7. Details of the dynamic process during adding and removing individual solitons.** (**A**) The zoom-in of the dynamic process (marked in Fig.3F by white arrow) in which a newly-appeared soliton changed a double-soliton unit into a triple-soliton unit. (**B**) The zoom-in of the dynamic process of a double-soliton unit (marked in Fig.3H by white arrow), in which $2^{nd}$-soliton was removed and it then becomes a single-soliton unit.

In this example, the change of pump power actually affected homogeneously upon each single soliton (due to the long response time of the EDFA). However, due to the noise perturbation, such abrupt change would only push some "unlucky" solitons beyond their threshold, which means that these soliton happened to be only slightly more vulnerable than the others at the beginning of the process. They might happen to be slightly deviated a bit more from the equilibrium positions than the others, or happen to have lower intensities. Under stable laser operations, such noise can be well compressed due to the presence of trapping potentials. Once the abrupt change of the laser pump power is strong enough, some "unlucky" solitons will be strongly affected and then disappear after a few hundreds of cavity round-trips. The remaining solitons, on the other hand, would successfully adjust their energies and group velocities according to the new working point, forming a stable supramolecular structure with a new pattern.

A significant difference between the adding and removing process of individual solitons lies in the duration of the dynamic process. While the removing of solitons starts immediately after the drop of pump power (as shown in Fig.3H), the adding of solitons usually requires a relatively long "preparation" time (around 8000 roundtrips as shown in Fig.3F) after the pump power has increased, since the newly-appeared solitons actually originated from the laser noise background, and needed to go through a complicated dissipative and nonlinear process before it could eventually acquire a soliton profile (*38*).

In the experiments we have also observed that the newly-added solitons could occasionally appear away from the "correct" positions when they initially emerged. Nevertheless, due to the presence of the long-range forces as depicted in Fig.1, E–G, these solitons would be quickly dragged backed to the "correct" (equilibrium) positions, and the supramolecular could still stably switch to a new pattern.